# Identifying Communities and Key Vertices by Reconstructing Networks from Samples

**Bowen Yan[1*], Steve Gregory[1]**
**1** Department of Computer Science, University of Bristol, Bristol BS8 1UB, UK
* E-mail: yan@cs.bris.ac.uk

## Abstract

Sampling techniques such as Respondent-Driven Sampling (RDS) are widely used in epidemiology to sample "hidden" populations, such that properties of the network can be deduced from the sample. We consider how similar techniques can be designed that allow the discovery of the structure, especially the community structure, of networks. Our method involves collecting samples of a network by random walks and reconstructing the network by probabilistically coalescing vertices, using vertex attributes to determine the probabilities. Even though our method can only approximately reconstruct a part of the original network, it can recover its community structure relatively well. Moreover, it can find the key vertices which, when immunized, can effectively reduce the spread of an infection through the original network.

## Introduction

The representation of data as networks has become increasingly popular in recent years in diverse application areas, including social science, biology, and epidemiology. However, real-world datasets are often incomplete because it may be difficult or impossible to collect complete data with available time or resources. In the case of network data, techniques have been developed to estimate properties of a network from samples of it, and various methods devised to discover or predict missing information, which can then be added to the network. For example, there are numerous strategies for finding missing edges [1] and even for finding the correct sign or direction of existing edges, in the context of signed [2] and directed [3] networks, respectively.

In contrast to the "edge prediction" problem, we are concerned with a different form of incomplete network. We assume that all edges, and vertices, are present in the network, but that vertices may have incorrectly been duplicated. This kind of error is very common, for example, in bibliographic databases and in genealogy, where a single individual may be known by more than one name and therefore represented by more than one vertex. Our task is to take a network with possibly duplicated vertices and attempt to *coalesce* all duplicates into a single vertex. Clearly, this can be done perfectly only if the identity of each vertex is known. However, if we have only partial information about vertices, we can still estimate the probability that each pair of vertices represent the same individual, and therefore predict which pairs should be coalesced. Having reconstructed the network, we can then attempt to identify features of it, such as the communities or the vertices which are important in some sense.

One potential application of this is epidemiology. It is well known that the structure of a contact network strongly influences the spread of an infectious disease in it. For example, in community-structured networks, infections travel fast within communities but slowly between them [4, 5]. High-degree vertices also play an important role in spreading disease [6]: it is effective to control disease transmission by immunizing high-degree people in a disassortative social network. There are some other practical applications, for example, identifying key individuals in a criminal network [7].

Knowledge of the network structure could clearly be invaluable in disease control or crime prevention, etc. The problem is that it is often very difficult to discover the structure of real networks, especially social networks. However, it is often possible to obtain samples (subgraphs) of the network. In epidemiology, techniques such as Respondent-Driven Sampling (RDS) [8] and contact tracing (CT) [9] are often used for this purpose. Our idea is to reconstruct networks by collecting samples based on paths (sequences of adjacent vertices) starting from vertices that are randomly selected from the population, and using these to reconstruct the network.

Our technique works as follows. First, a member of the population is randomly selected as a "seed", which begins the first *sample path*. Then one of the neighbors of the seed is randomly selected and added to the end of the path. The process is repeated from the last vertex that was added to the path. If the selected neighbor already exists in any of the paths collected so far, we choose a different neighbor. The path stops growing if the latest vertex has no neighbors that have not already been sampled. When a path stops growing, a new path is started, beginning at a new randomly-selected seed. This is repeated until the total size (number of vertices) of all paths reaches a specified limit. We call this the Random Path Method (RPM). In our High-degree Path Method (HPM), we choose the highest-degree neighbor at each step, instead of a random one.

Both methods produce a set of paths: linear samples of the network about which we have complete information. In particular, the sampling strategy implies that we know the identity of every vertex in the paths, in order to avoid duplicates. For example, in the case of social networks, the members of these paths can be interviewed to obtain arbitrarily detailed information about them. In the second stage of our process, we convert each sample path to a *sample tree*, in which the seed becomes the root of the tree and the successor of each vertex $v$ in the path becomes an offspring of $v$ in the tree. Now a vertex $v$ may have additional offspring in the tree: these are neighbors of $v$ in the underlying network, and may (but need not) include $v$'s predecessor and successor in the path. In the case of social networks, these additional offspring of $v$ are "friends" of $v$ which have been reported by $v$, but we do not have complete knowledge about them; in particular, they may be duplicated. Thus, a sample tree contains vertices of two types, which we call *respondents* and *friends*.

Each respondent is asked to provide a description of each chosen friend which, for practical reasons, is very unlikely to uniquely identify the friend. We use these descriptions to choose which pairs of vertices might represent the same individual, and coalesce them into a single vertex. Naturally, the success of this technique depends on the accuracy of descriptions (increased detail reduces the chance of incorrect matching) as well as on the size and number of the trees. Only if the trees cover the whole network and the descriptions are unique can we expect to reconstruct the underlying network perfectly. However, by exploiting knowledge of the frequency of individuals in the population matching each description, we can probabilistically coalesce vertices to build a large ensemble of possible networks, from which we can estimate some properties of the underlying network.

We explore this technique in the remainder of the paper. Essentially, we evaluate the similarity between the reconstructed network and the original underlying network, varying the features of the underlying network and the parameters of the sampling procedure. "Similarity" can be defined in several ways: similarity of edges, similarity of community structure, and similar ranking of various vertex properties. Finally, we use the technique to identify key vertices in a contact network for an infectious disease (i.e., those that might be important to the spread of infection) and immunize them. We find that choosing the key vertices from a relatively small sample can control the spread of infection almost as well as choosing them from the entire network.

## Methods

### *The underlying network*

We assume the existence of an *a priori* unknown network, which we call the *underlying network*. This is a simple undirected, unweighted network. Every vertex is characterized by a single, discrete-valued, attribute. For example, in a social network, this could represent any real-life property of the individual, such as gender, age, hair color, etc., or a tuple of them, e.g., (male, 34, brown). In general, this divides the vertices into a number ($g$) of categories. Therefore, without loss of generality, we assume that the attribute of $v$, $a_v$, is an integer in the range $[1,g]$. The distribution of attribute values can vary; for example, a uniform distribution could be used for age while a normal distribution is more realistic for height.

### *Sampling procedure*

The first phase of our procedure collects a sample comprising a set of paths with a total size of approximately $n_r$ vertices:

1. Choose any vertex $s$ (that does not already appear in a path) from the underlying network, as seed.
2. Add $s$ to a new path.
3. If size of all paths = $n_r$, terminate.
4. For the most recently added vertex $v$, try to obtain a neighbor $u$ of $v$ that is not in any path.
5. If $u$ is found, add $u$ to current path and repeat from step 3.
6. If $u$ is not found, repeat from step 1.

In step 4, the vertex that we find must be a neighbor of $v$ in the underlying network and must not have already been added to a path. This way we construct paths sequentially, until the total size is approximately the required size, $n_r$.

In the second phase, we obtain from each respondent vertex $v$ up to $f$ "friends", which again must be neighbors of $v$ in the underlying network. If $v$ has more than $f$ neighbors, any $f$ of the neighbors can be chosen. If $v$ has $f$ neighbors or fewer, all of $v$'s neighbors are chosen. The friends chosen by $v$ are added as offspring of $v$ in the *sample tree*. A vertex can be chosen as a friend even if it already appears in a tree. Finally, each friend vertex $u$ is labeled with a "description" $d_u$: a set of categories that includes $u$'s actual category. Descriptions are assumed to be correct but not necessarily precise. For example, if the vertex has (age) attribute 34 in the underlying network, its description in the sample tree could be, for example, {33,34} or {34,35,36}. If the same network vertex appears as a friend more than once, each occurrence may have a different (correct) description. The final trees contain $n_r+n_f$ vertices: $n_r$ respondents and $n_f$ friends.

Figure 1(b) shows one possible set of sample trees, sampled from the underlying network of Figure 1(a), with vertices 1 and 3 as seeds. Notice that a vertex (e.g., 31) appears more than once in the trees (e.g., 31, 31', 31"), if it is a friend of more than one respondent.

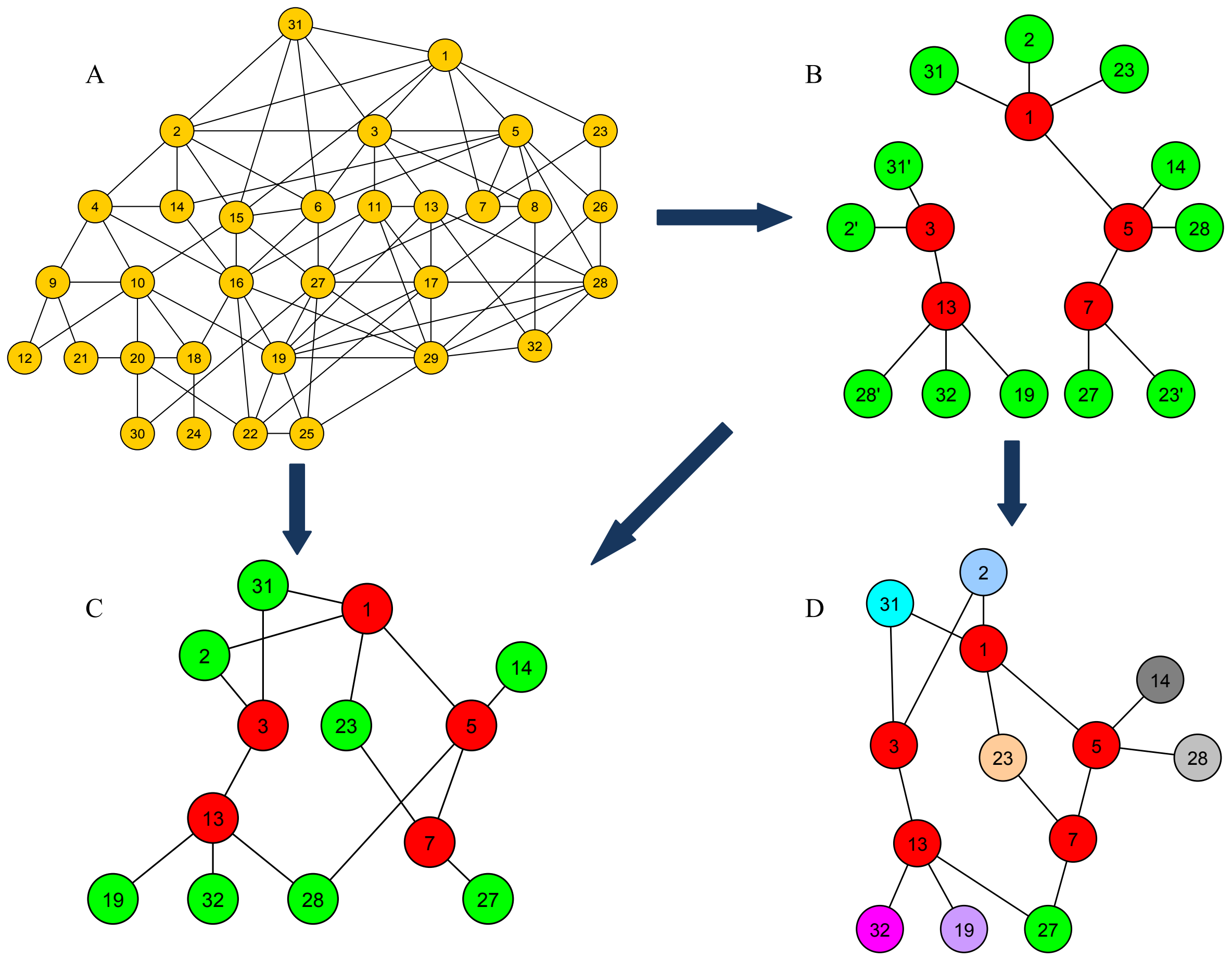

**Figure 1. Underlying network, samples, and reconstructed network.** A. Underlying network. B. Sample paths, with respondents in red and friends in green. C. True network. D. Reconstructed network: green vertices represent incorrectly coalesced pairs.

### *Network reconstruction*

In this phase we attempt to convert the sample trees to a network by coalescing all vertices that represent the same vertex in the underlying network. If we could do this perfectly, we would end up with a subgraph of the underlying network which contains all nodes and edges that appear in the sample trees; we call this subgraph the *true network*. Figure 1(c) shows the true network corresponding to Figure 1(b).

Our network reconstruction algorithm repeatedly performs the following steps:

1. Choose any two vertices $u$ and $v$ from the sample trees.
2. Calculate the probability $p$ that $u$ and $v$ represent the same individual.
3. With probability $p$, coalesce $u$ and $v$.
4. If $u$ is a respondent and $v$ is a friend, the coalesced vertex is $u$, still labeled with $a_u$.
5. If $u$ and $v$ are both friend vertices, the coalesced vertex is a friend vertex labeled with the description $d_u \cap d_v$.

These steps are repeated until the network size is reduced from $n_r$+$n_f$ vertices to the desired size, $n_t$.

The probability $p$ calculated in step 2 depends on whether $u$ and $v$ are respondents or friends. If they are both respondents, $p$=0 (all respondents are known to be distinct).

For respondent $u$ and friend $v$:

- If there is an edge $\{u,v\}$, $p$=0, because a respondent cannot choose himself as a friend.
- If $a_u \notin d_v$ ($u$ does not satisfy the description of $v$), $p$=0.
- Otherwise,

$$p = \frac{1}{n_t \Pr(d_v)}. \quad (1)$$

where $\Pr(d_v)$ is the probability of an arbitrary vertex satisfying description $d_v$. This is because there are, on average, $n_t \Pr(d_v)$ vertices satisfying that description and $u$ is equally likely to be any of them.

For friend $u$ and friend $v$:

- If $u$ and $v$ have a common neighbor (a respondent), $p$=0, because all friends of a respondent are distinct.
- Otherwise,

$$p = \frac{(\Pr(d_u \cap d_v)/\Pr(d_u))(\Pr(d_u \cap d_v)/\Pr(d_v))}{n_t \Pr(d_u \cap d_v)} = \frac{\Pr(d_u \cap d_v)}{n_t \Pr(d_u)\Pr(d_v)}. \quad (2)$$

The probabilities $\Pr(d)$ depend on what is known about the underlying network. For example, if descriptions are ages (in years) and are uniformly distributed over a range of 50 years, Pr({34,35,36}) = 3/50. If we try to match two friend vertices with descriptions {33,34} and {34,35,36}, $p$ would be $(1/50)/(n_t(2/50)(3/50)) = 50/6n_t$, assuming that $n_t$ is reasonably large.

Figure 1(d) shows one possible network produced by our network reconstruction. Most duplicate vertices have been coalesced correctly ({2,2'}, {23,23'}, and {31,31'}), but one pair ({27,28'}) has been merged wrongly, shown in green.

### *Experimental framework*

In the next section, we evaluate our method empirically. In order to evaluate our reconstructed networks, we begin with a known underlying network and simulate our sampling procedure on it.

For the underlying network we use a selection of real social networks, listed in Table 1. We also use synthetic benchmark networks generated by the method of Lancichinetti et al. [10], which allows us to vary the parameters, especially the strength of community structure. The LFR networks have several parameters:

1. $n$ is the number of vertices.
2. $\langle k \rangle$ and $k_{max}$ are the average and maximum degree.
3. $\tau_1$ and $\tau_2$ are the exponents of the power-law distribution of vertex degrees and community sizes.
4. $c_{min}$ and $c_{max}$ are the minimum and maximum community size.
5. $\mu$ is the mixing parameter: each vertex shares, on average, a fraction $\mu$ of its edges with vertices in other communities.

Having obtained or created the network, we then label the vertices with attributes which are integers in [1,$g$]; the number of categories, $g$, can be varied. The values are assigned randomly, according to a uniform or normal distribution. As an optional step, the network can then be made more assortative by swapping the attribute values of two vertices so that the values of neighboring vertices become more similar. By default, for most experiments in this paper, we use a normal distribution and do not try to increase assortativity.

**Table 1.** Real-world networks

| Networks | Ref. | Type | Vertices | Edges |
|---|---|---|---|---|
| Email | [11] | social | 1133 | 5451 |
| Blogs | [12] | social | 3982 | 6803 |
| CA-GrQc | [13] | collaboration | 5242 | 28980 |
| Erdös1997 | [14] | collaboration | 5482 | 8972 |

In our Random Path Method, each seed is chosen randomly from the underlying network and each subsequent vertex is chosen randomly from the neighbors. In our High-degree Path Method, each seed is randomly chosen from vertices with degree 5 or more, and we always add the highest-degree neighbor to the path. The friends (at most $f$) of $v$ are selected randomly from $v$'s neighbors in the underlying network. The description attached to each friend is a randomly-chosen range of $c$ (which can be varied) consecutive categories that includes the actual category of the corresponding vertex in the underlying network.

In our experiments, we assume that we know the size of the true network, $n_t$, so that the network reconstruction phase stops coalescing vertices when that size is reached. This allows us to take the true network as the best possible reconstructed network. In real use, $n_t$ would not be known, but it should be possible to devise suitable alternative termination criteria; for example, to stop coalescing when the average degree (or other network property) of the reconstructed network reaches a realistic value.

## Results and Discussion

### *Evaluation of the algorithm*

First we evaluate our method for reconstructing networks: in particular, how its accuracy is affected by the properties of the underlying network and the parameters of the sampling procedure. Network properties are:

1. Number of categories of vertex attributes, $g$.

2. Distribution of vertex attributes: e.g., uniform or normal distribution.
3. Whether the vertex attributes are assortative or non-assortative. In an assortative network we attempt to make vertex attributes similar to their neighbors', while attributes in the non-assortative network are randomly assigned.
4. Strength of community structure, $\mu$: each vertex shares, on average, a fraction $\mu$ of its edges with vertices in other communities.

By default, we use normally distributed, non-assortative, networks. Parameters of the sampling procedure include:

1. Number of neighbors named as friends, $f$.
2. Number of consecutive categories in descriptions of friends, $c$. Increasing $c$ reduces the precision of the description; for example, describing a friend's age to the nearest 5 years ($c$=5) instead of giving the exact age in years ($c$=1).
3. Size of reconstructed network after coalescing, $n_t$.

Default values that we use are $f$=5 (a common value in surveys of friendship networks) and $c$=1, while $n_t$ is set to 8% of the underlying network size; this is approximately the fraction used in some real surveys of "hidden populations" such as drug users. However, we also examine the effect of varying each of these parameters.

To measure the accuracy of the network reconstruction, we use *coalescing precision*, which we define as:

$$\text{coalescing precision} = \text{number of correct coalesced pairs/number of coalesced pairs.} \tag{3}$$

where a pair of vertices $\{u,v\}$ in the sample trees is a *coalesced pair* if our method coalesces them, and a *correct coalesced pair* if (also) $u$ and $v$ represent the same vertex in the underlying network.

Figure 2 shows the effect of varying $g$ on coalescing precision, for both assortative and non-assortative networks. A high $g$ represents more accurate, less ambiguous, descriptions. As expected, the more information about descriptions obtained, the higher the precision. Our method performs better on non-assortative networks than assortative networks because, in the latter, the descriptions of vertices in sample trees are more likely to be similar and therefore harder to coalesce correctly. The High-degree Path Method performs better than the Random Path Method, and this is true of subsequent experiments; we discuss this in the conclusions.

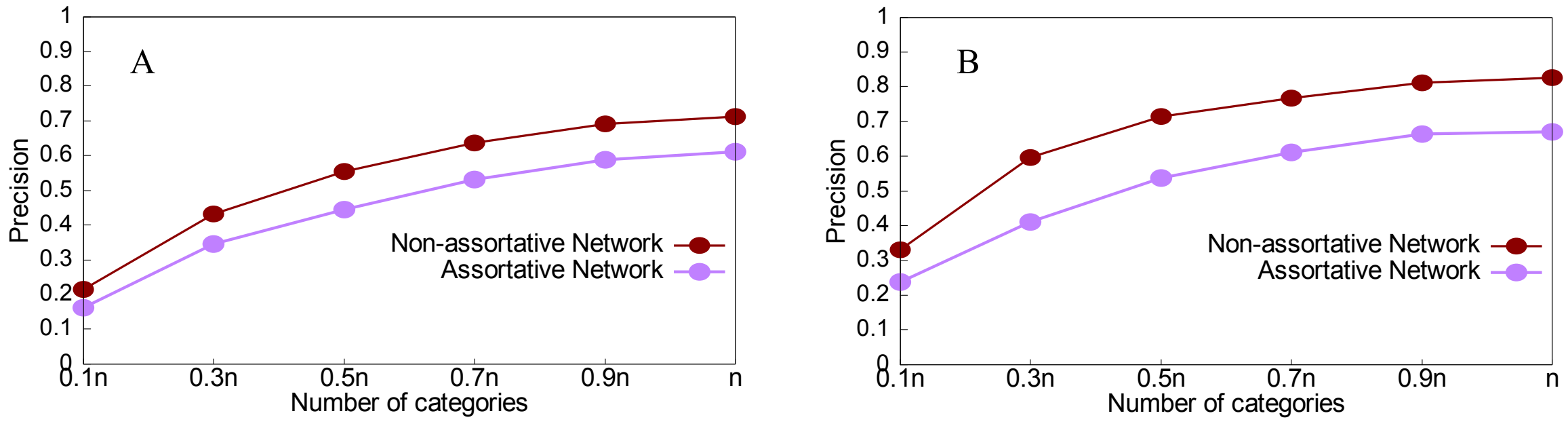

**Figure 2. Performance of our method.** Precision, for varying $g$, on assortative and non-assortative LFR networks with normal distribution with $n$=1460, $\langle k \rangle$=20, $k_{max}$=30, $\mu$=0.1, $\tau_1$=3, $\tau_2$=1, $c_{min}$=10, and $c_{max}$=20. A. Random Path Method. B. High-degree Path Method.

Figure 3 shows the effect of different distributions of vertex attributes for the underlying networks on coalescing precision. Our method works slightly better for uniform than for normal distribution; this is because, with a normal distribution, more individuals fall into a smaller number of categories, so the categories are less effective in distinguishing them.

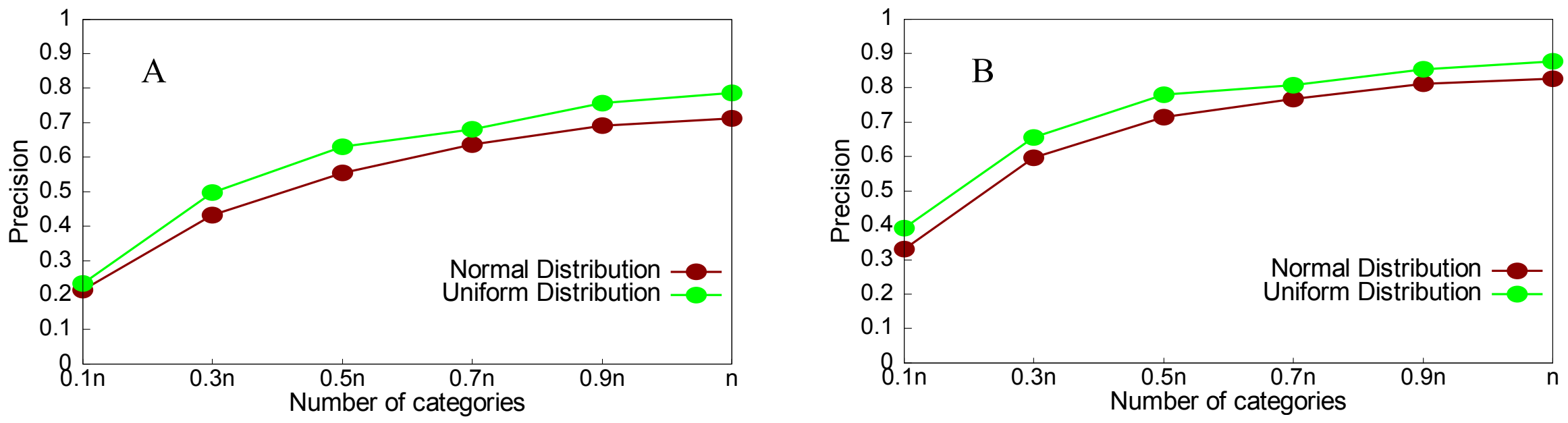

**Figure 3. Performance of our method.** Precision, for normal distribution and uniform distribution of vertex attributes, on non-assortative LFR networks with $n$=1460, $\langle k \rangle$=20, $k_{max}$=30, $\mu$=0.1, $\tau_1$=3, $\tau_2$=1, $c_{min}$=10, and $c_{max}$=20. A. Random Path Method. B. High-degree Path Method.

Figure 4 shows the effect of community structure on coalescing precision. The parameter $\mu$ controls the community structure of the underlying network. When we increase $\mu$ (for a less distinct community structure), the coalescing precision gradually falls. This is because, with a smaller $\mu$, the vertices in each sample tree are more likely to belong to the same community and therefore any coalescing of them is more likely to be correct by chance.

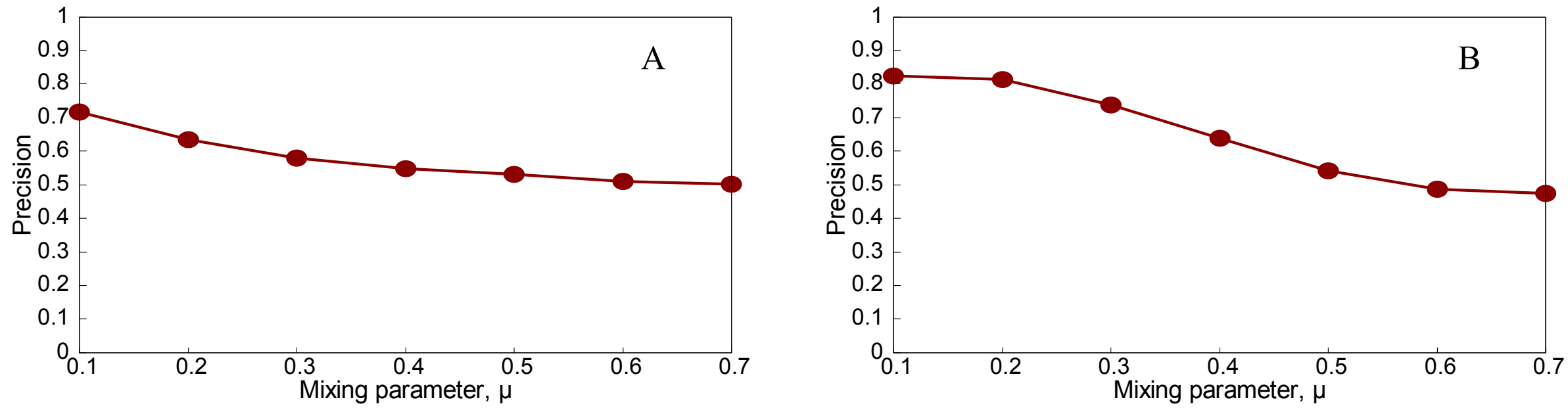

**Figure 4. Performance of our method.** Precision, varying $\mu$ on non-assortative LFR networks with normal distribution with $n$=1460, $\langle k\rangle$=20, $k_{max}$=30, $\tau_1$=3, $\tau_2$=1, $c_{min}$=10, $c_{max}$=20, and $g$=$n$. A. Random Path Method. B. High-degree Path Method.

As $f$ increases, the sample trees tend to become larger, and so we need fewer of them because $n_t$ (the total size of all trees) is fixed. They also contain a larger fraction of friends, which carry less information than respondents and, unlike them, might be duplicated. Therefore, we expect worse results when $f$ is larger. Figure 5 shows the coalescing precision when we vary $f$. In this example, the number of friends does not noticeably affect the coalescing precision of the method. Therefore, we use $f$=5, because it is a common value in surveys of friendship.

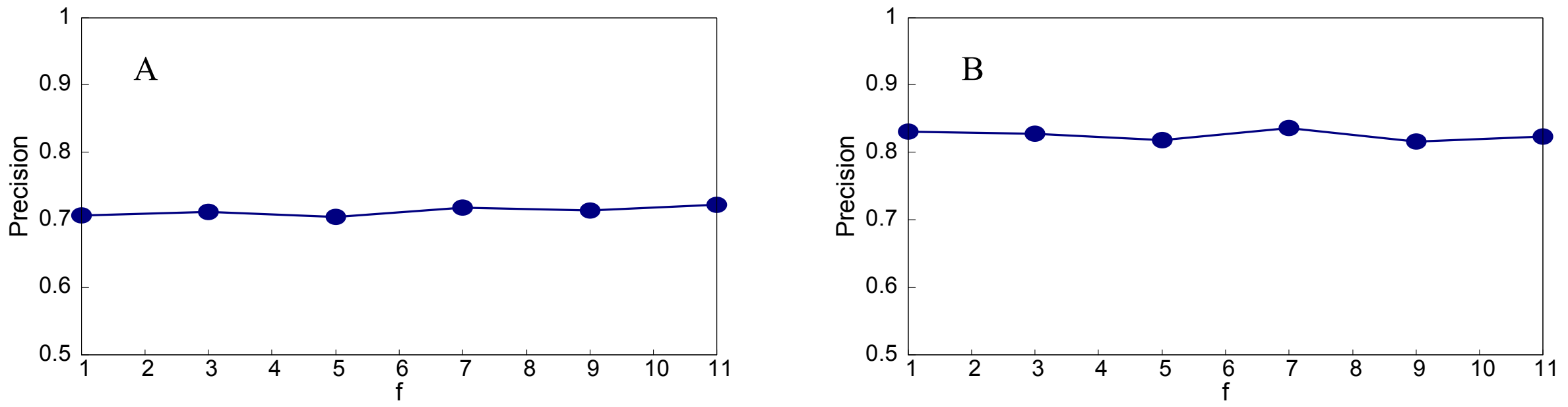

**Figure 5. Performance of our method.** Precision, varying $f$ on non-assortative LFR networks with normal distribution with $n$=1460, $\langle k\rangle$=20, $k_{max}$=30, $\mu$=0.1, $\tau_1$=3, $\tau_2$=1, $c_{min}$=10, $c_{max}$=20, and $g$=$n$. A. Random Path Method. B. High-degree Path Method.

In reality, the descriptions of friends are not necessarily accurate, for example, a range of heights can be described for a friend. Therefore, we vary $c$, the number of categories in a range for describing friends. Figure 6 shows the effect of different ranges of categories on coalescing precision. As expected, a wider range results in lower precision because there are more pairs of vertices that could be (incorrectly) coalesced.

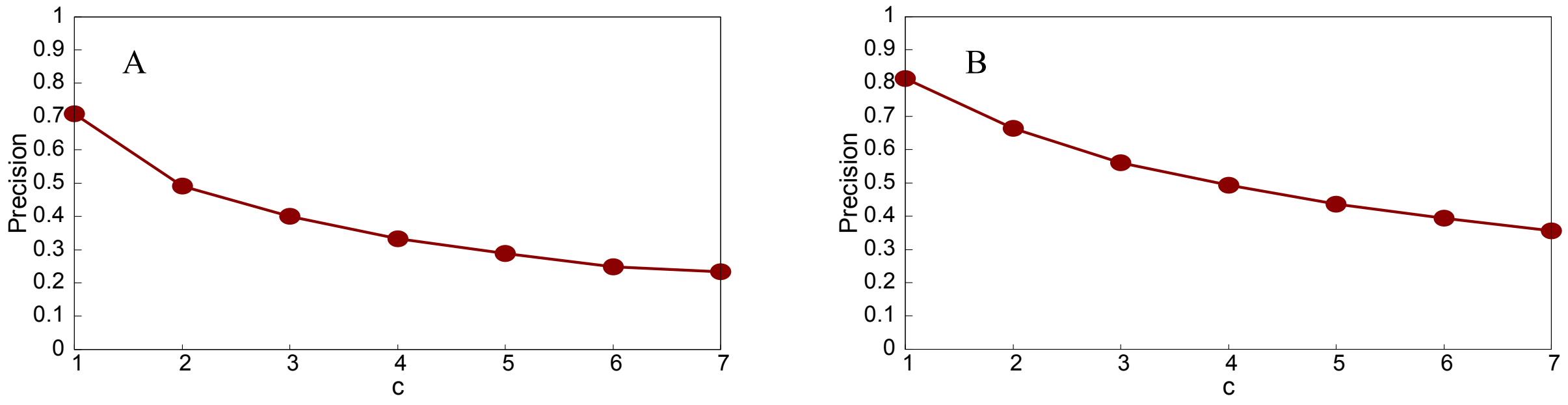

**Figure 6. Performance of our method.** Precision, varying $c$ on non-assortative LFR networks with normal distribution, with $n$=1460, $\langle k\rangle$=20, $k_{max}$=30, $\mu$=0.1, $\tau_1$=3, $\tau_2$=1, $c_{min}$=10, $c_{max}$=20, and $g$=$n$. A. Random Path Method. B. High-degree Path Method.

*Community structure detection*

The topological features of networks have attracted much attention in recent disease transmission research [5]. In particular, in a contact network, a disease spreads quickly between individuals inside communities and slowly between communities. Our hypothesis is that community structure in the underlying network can be predicted by the community structure found in the reconstructed network. We test this by using a good community detection algorithm (Infomap [15]) to detect communities in the reconstructed and true networks and then using the Normalized Mutual Information (NMI) measure to compare the partitions found in each network. However, we cannot use this to compare the partitions obtained in the underlying network and reconstructed network, because the reconstructed network omits most of the vertices and edges from the underlying network. Therefore, we use another measure, which we call *community precision*:

$$\text{community precision} = \text{number of pairs in same community in reconstructed and underlying networks} / \text{number of pairs in the same community in the reconstructed network.} \quad (4)$$

Figure 7 shows a comparison of the partitions obtained in the true network and the reconstructed network by using NMI, and varying $g$. Figure 8 shows the community precision between the underlying and reconstructed networks. Both results show substantial similarity between the correct communities and those found in the reconstructed network. Figure 7 shows that community structure is found more accurately when the network is assortative even though the coalescing precision is lower (Figure 2). This is because, with our reconstruction method, even incorrect coalescing is likely to coalesce vertices that are in the same community if the network is assortative.

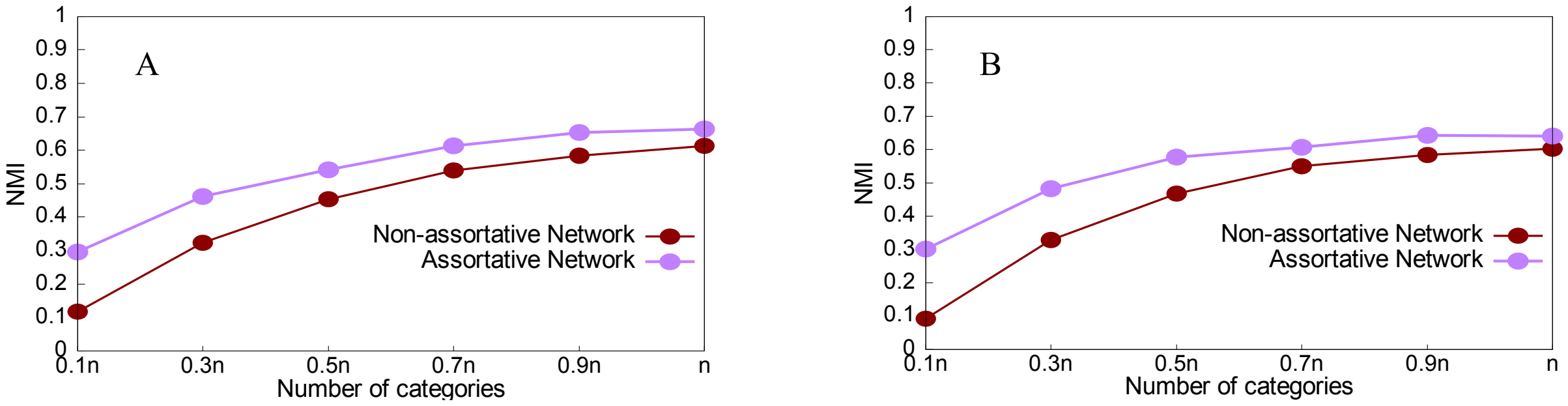

**Figure 7. Comparison of partitions obtained in true network and reconstructed network.** Normalized Mutual Information, varying $g$, with $n$=1460, $\langle k \rangle$=20, $k_{max}$=30, $\mu$=0.1, $\tau_1$=3, $\tau_2$=1, $c_{min}$=10, and $c_{max}$=20. A. Random Path Method. B. High-degree Path Method.

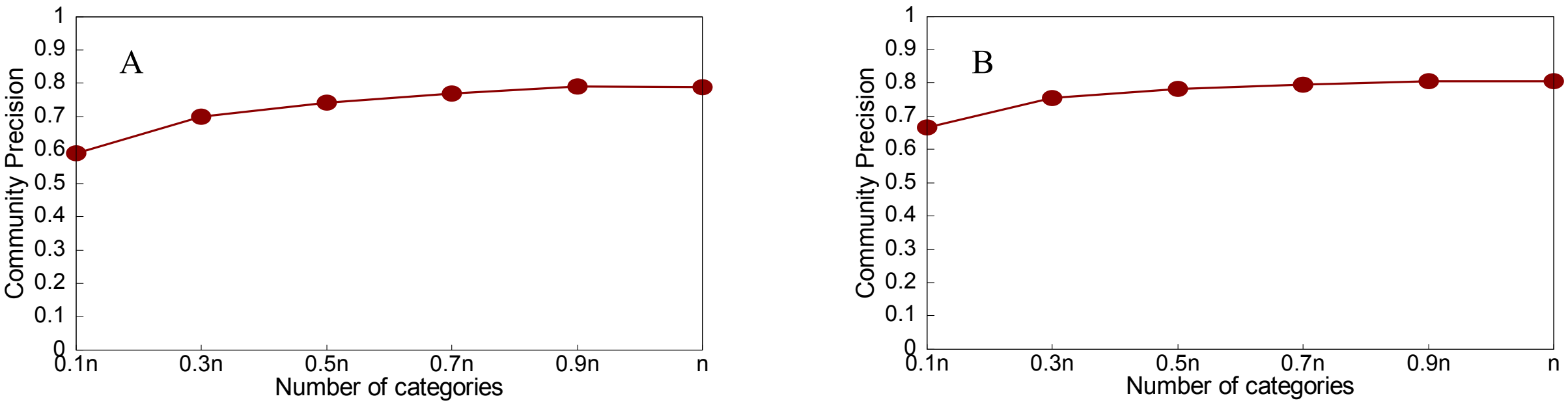

**Figure 8. Comparison of partitions obtained in underlying network and reconstructed network.** Community precision, varying $g$, with $n$=1460, $\langle k \rangle$=20, $k_{max}$=30, $\mu$=0.1, $\tau_1$=3, $\tau_2$=1, $c_{min}$=10, and $c_{max}$=20. A. Random Path Method. B. High-degree Path Method.

*Rank correlation*

In addition to detecting community structure, we are also interested in identifying the important vertices, which may be more important in spreading disease. Controlling (e.g., immunizing) key vertices could prevent a current or potential epidemic. In our experiments, we consider three vertex properties:

1. *degree*($v$): the number of neighbors of a vertex $v$. The vertices with high degree are important.

2. $k_{out}(v)$: the number of neighbors of a vertex $v$ in other communities. A vertex with a high $k_{out}$ is important because it is a "bridge" vertex.
3. *embeddedness*($v$): the fraction of neighbors of a vertex $v$ that belong to the same community as this vertex [16]. A "bridge" vertex should have a low embeddedness. Embeddedness is related to the other two properties by the equation:

$$embeddedness(v) = (degree(v) - k_{out}(v)) / degree(v). \quad (5)$$

We want to discover whether vertices in the reconstructed network that are "important" (according to each of these three properties) play the same role in the underlying network. The procedure follows the steps below:

1. Calculate different properties for each of the vertices in the reconstructed network and the underlying network separately (ignoring vertices that do not occur in the reconstructed network).
2. Rank the vertex lists according to those values.
3. Compare the ranked lists of the reconstructed network and the underlying network by using a rank correlation coefficient. Spearman's rank correlation coefficient [17] has been chosen: this measures the similarity between two ranked lists of the same set of items.

Figures 9-12 show the results for our four real networks. In all cases, the similarity between the true network (TN) and underlying network (ULN) is less than perfect, because the true network lacks many of the edges present in the underlying network. The similarity between the reconstructed network (RCN) and true network increases with $g$. The similarity between the reconstructed network and underlying network is lower still but also increases with $g$. Degree is slightly better predicted than $k_{out}$, while embeddedness is the hardest to predict of the three measures.

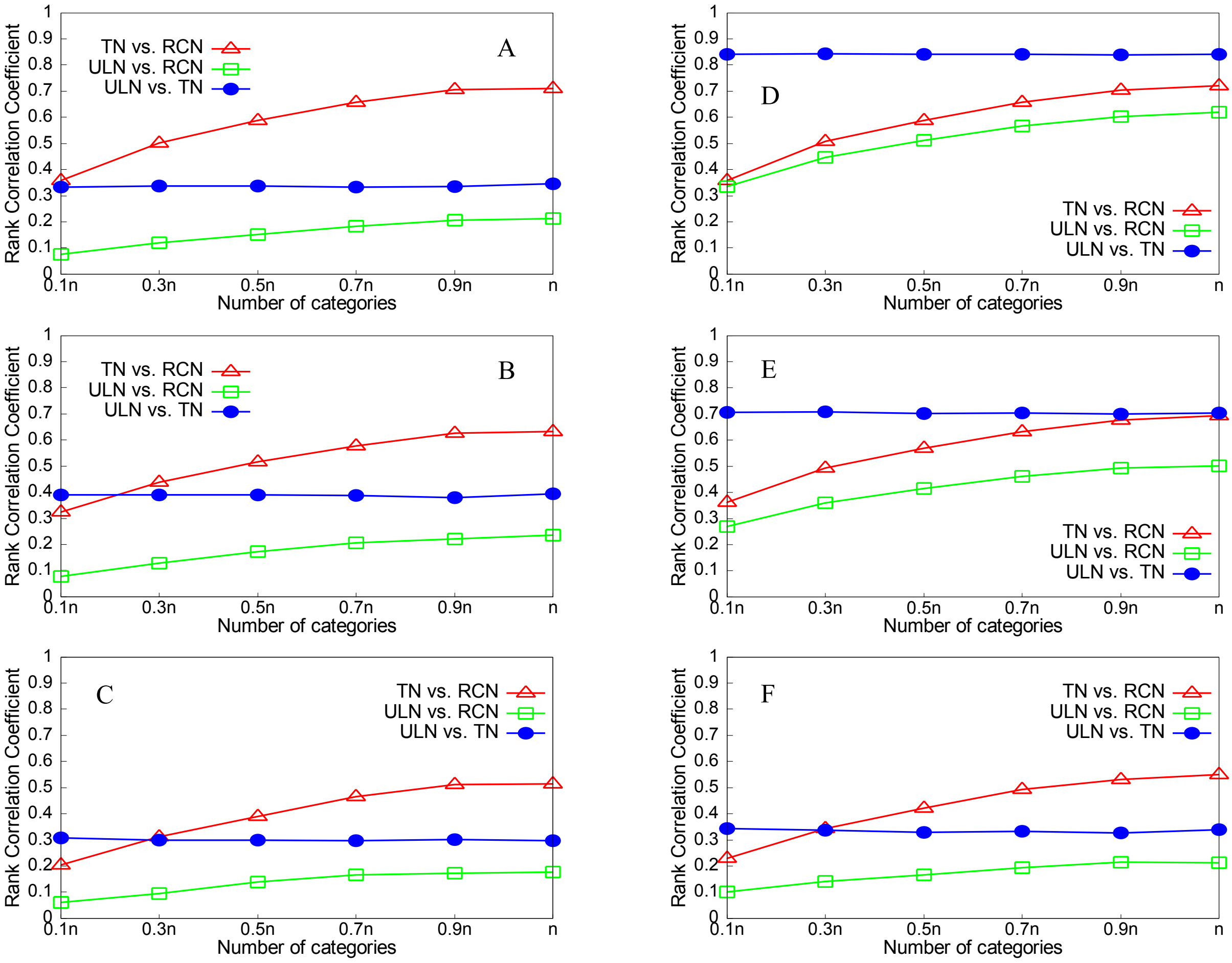

**Figure 9. Rank correlations for Email network.** A. Degree (RPM). B. $k_{out}$ (RPM). C. Embeddedness (RPM). D. Degree (HPM). E. $k_{out}$ (HPM). F. Embeddedness (HPM).

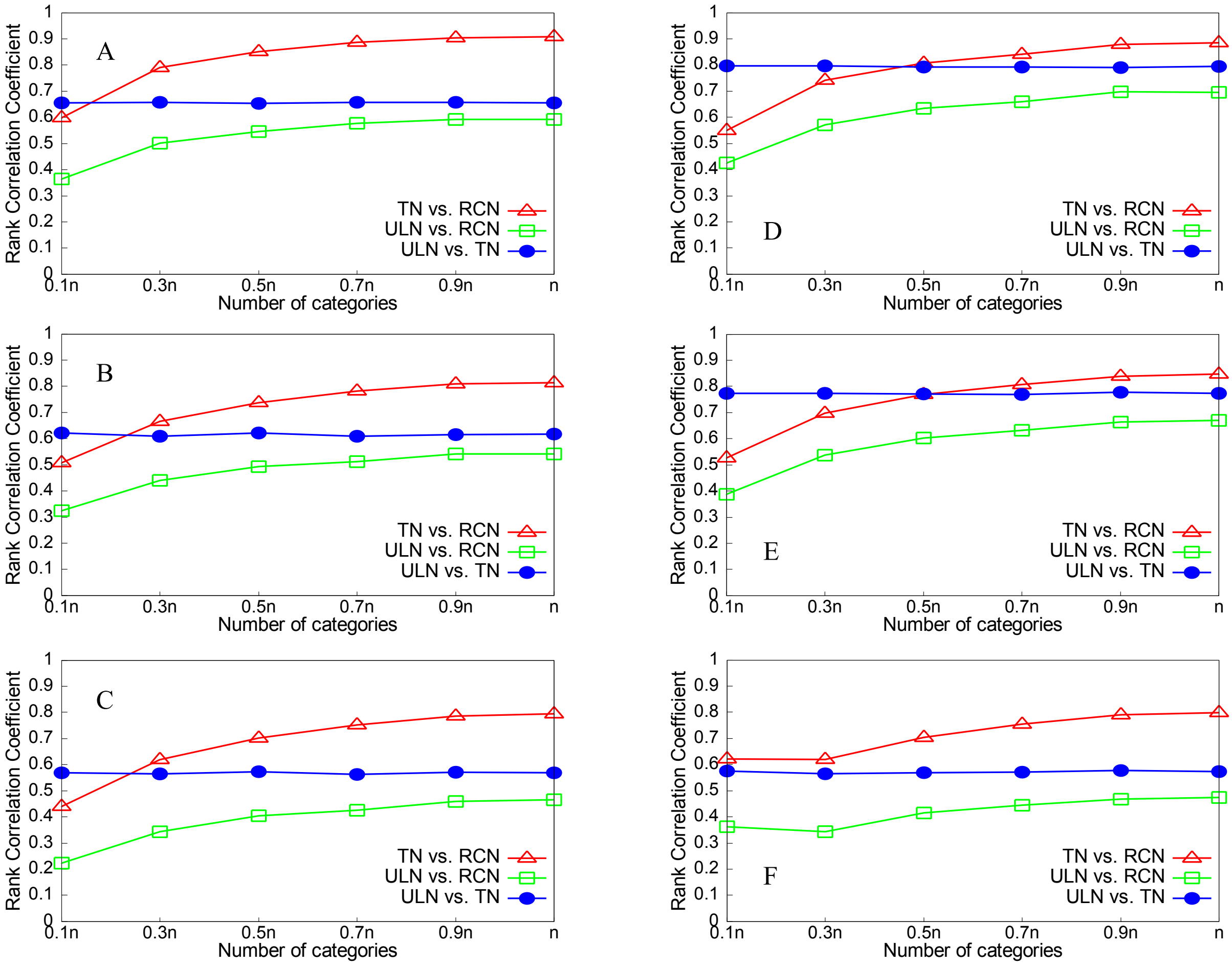

**Figure 10. Rank correlations for Blogs network.** A. Degree (RPM). B. $k_{\mathrm{out}}$ (RPM). C. Embeddedness (RPM). D. Degree (HPM). E. $k_{\mathrm{out}}$ (HPM). F. Embeddedness (HPM).

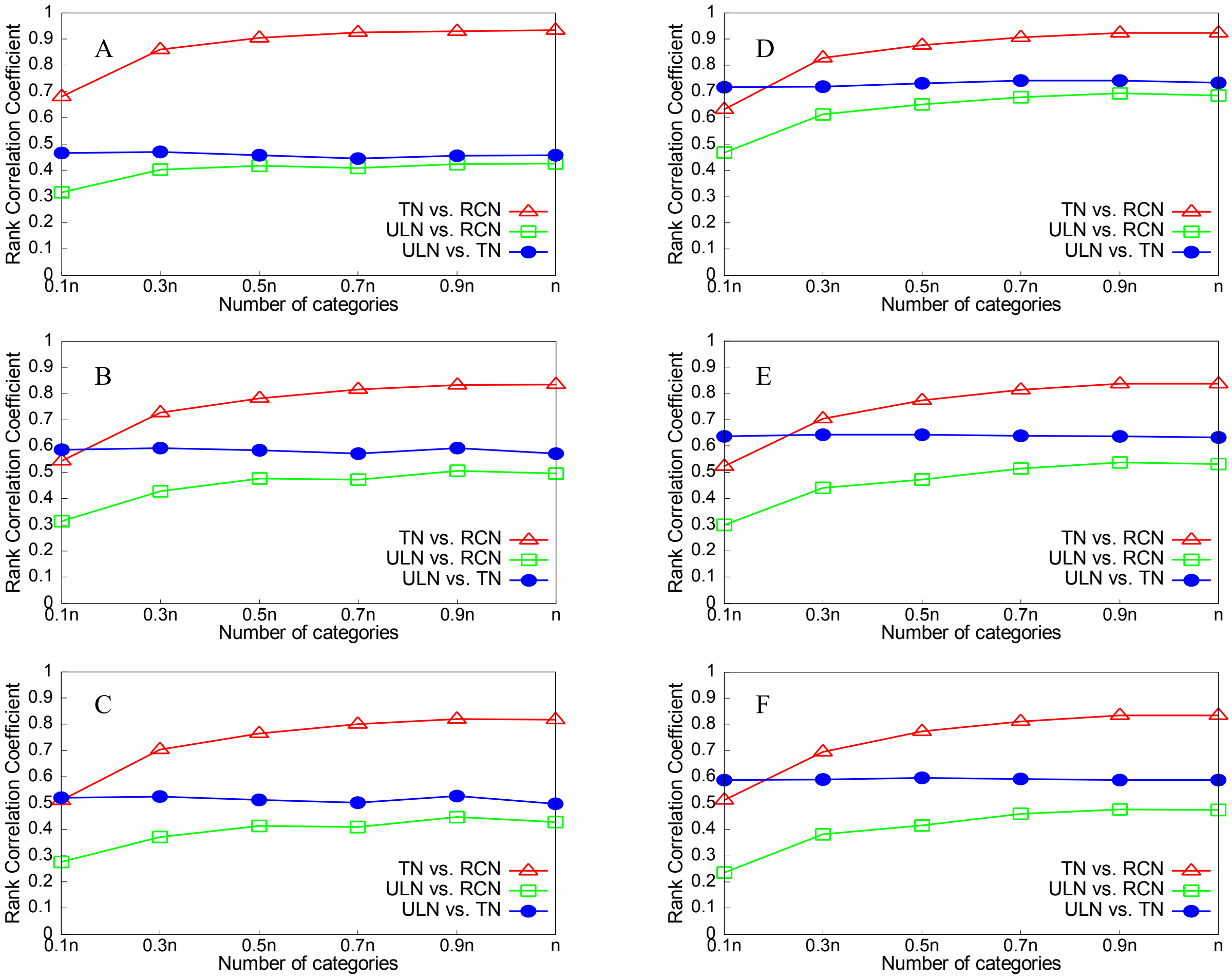

**Figure 11. Rank correlations for CA-GrQc network.** A. Degree (RPM). B. $k_{out}$ (RPM). C. Embeddedness (RPM). D. Degree (HPM). E. $k_{out}$ (HPM). F. Embeddedness (HPM).

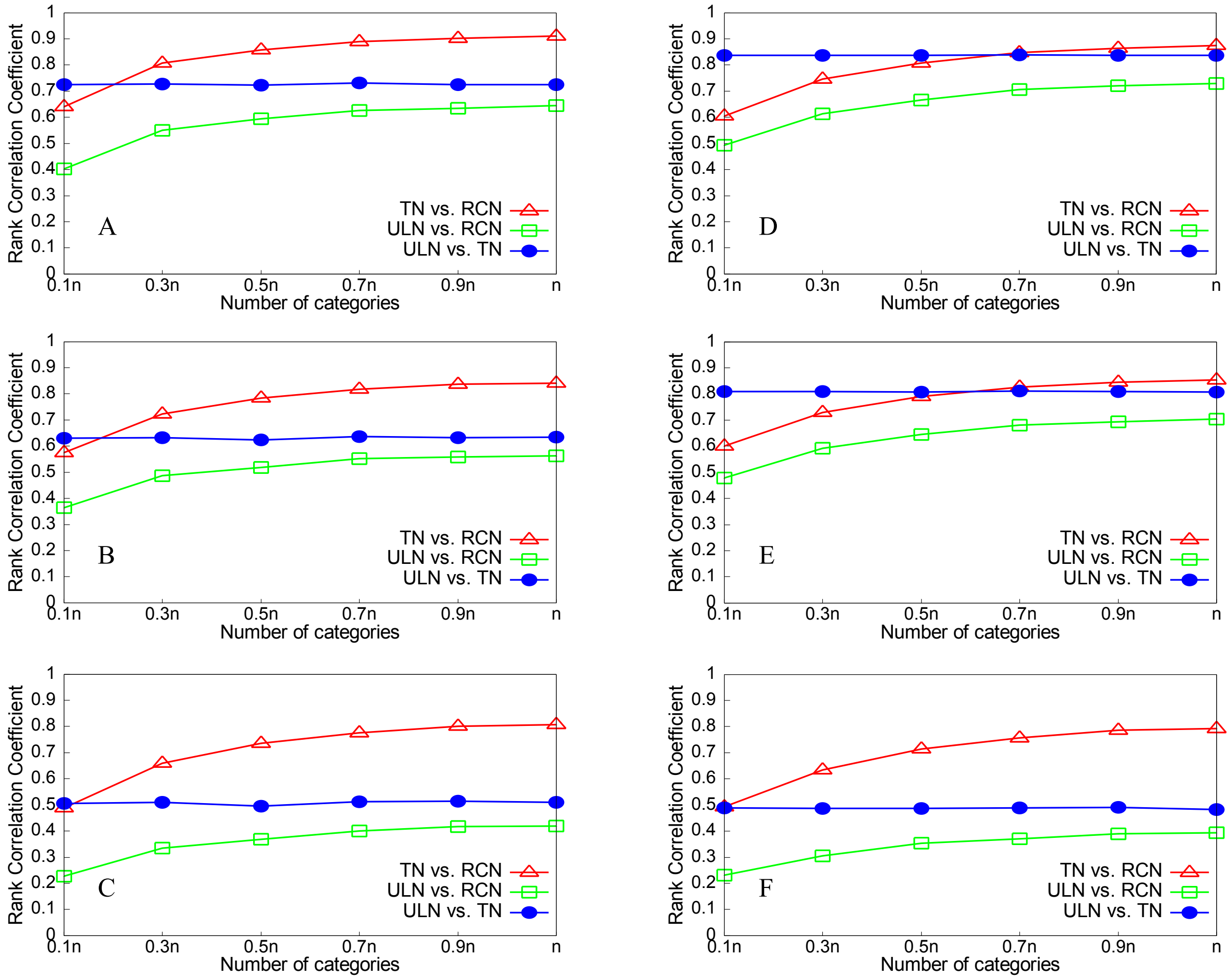

**Figure 12. Rank correlations for Erdös1997 network.** A. Degree (RPM). B. $k_{out}$ (RPM). C. Embeddedness (RPM). D. Degree (HPM). E. $k_{out}$ (HPM). F. Embeddedness (HPM).

### *Disease simulation*

In the last section, we showed that it is possible to predict certain vertex properties of a network from a sample, but only with moderate accuracy. Moreover, since the sample contains only a small fraction (8% in our experiments) of the underlying network, our method can predict nothing about the remaining 92% of vertices. The question is whether it is useful to have this imperfect information about the vertex properties of such a small sample of the network. We investigate this by using this information (the key vertices in the reconstructed network) to try to control the spread of disease in the underlying network.

The experiments are designed as follows:

1. Calculate different properties for vertices in the reconstructed network (8% of the underlying network size) and order the list of vertices. We calculate the average value of each vertex property over 100 instances of such networks.
2. Immunize the top-ranked vertices.
3. Simulate a disease outbreak in the underlying network.
4. Measure the total epidemic size, to assess the effectiveness of the control strategy.

We compare this method with an alternative whereby the actual value of each property is calculated using the whole underlying network. This is not feasible in real life because the underlying network is unknown, but can provide an upper bound for comparison with our strategies.

In step 3, we simulate the spread of an infectious disease using a susceptible-infected-recovered (SIR) model, taking the real-world networks in Table 1 as the underlying networks. We initially infect 0.2% of the vertices in the underlying network, then at each time step, there is a fixed probability (0.08) of the disease being transmitted to a susceptible vertex from each infectious neighbor, and a vertex remains infectious for a fixed period (four time steps).

Figures 13-16 (a) show results of the three strategies applied to the underlying network, together with a strategy of immunizing vertices selected randomly from the whole network. Figures 13-16 (b) and (c) show the same strategies where the vertex is chosen only from the reconstructed network with our two sampling methods. Because the latter is much smaller than the underlying network (8% of the size), the results are inevitably worse than choosing from the complete network, but still

remarkably good. It may be thought that the immunization strategy is influenced only by *which* vertices appear in the sample trees, and hence reconstructed networks, rather than the *values* of the vertex properties. For example, vertices in sample trees naturally have a higher degree than average. To show that this is not the case, Figures 13-16 (b) and (c) also show a "random" strategy in which vertices are randomly selected from the reconstructed network. Here, because each reconstructed network contains a different set of vertices, we order vertices by the frequency with which they appear in reconstructed networks. Although this performs better than random selection from the whole underlying network, it is less good than our proposed strategies.

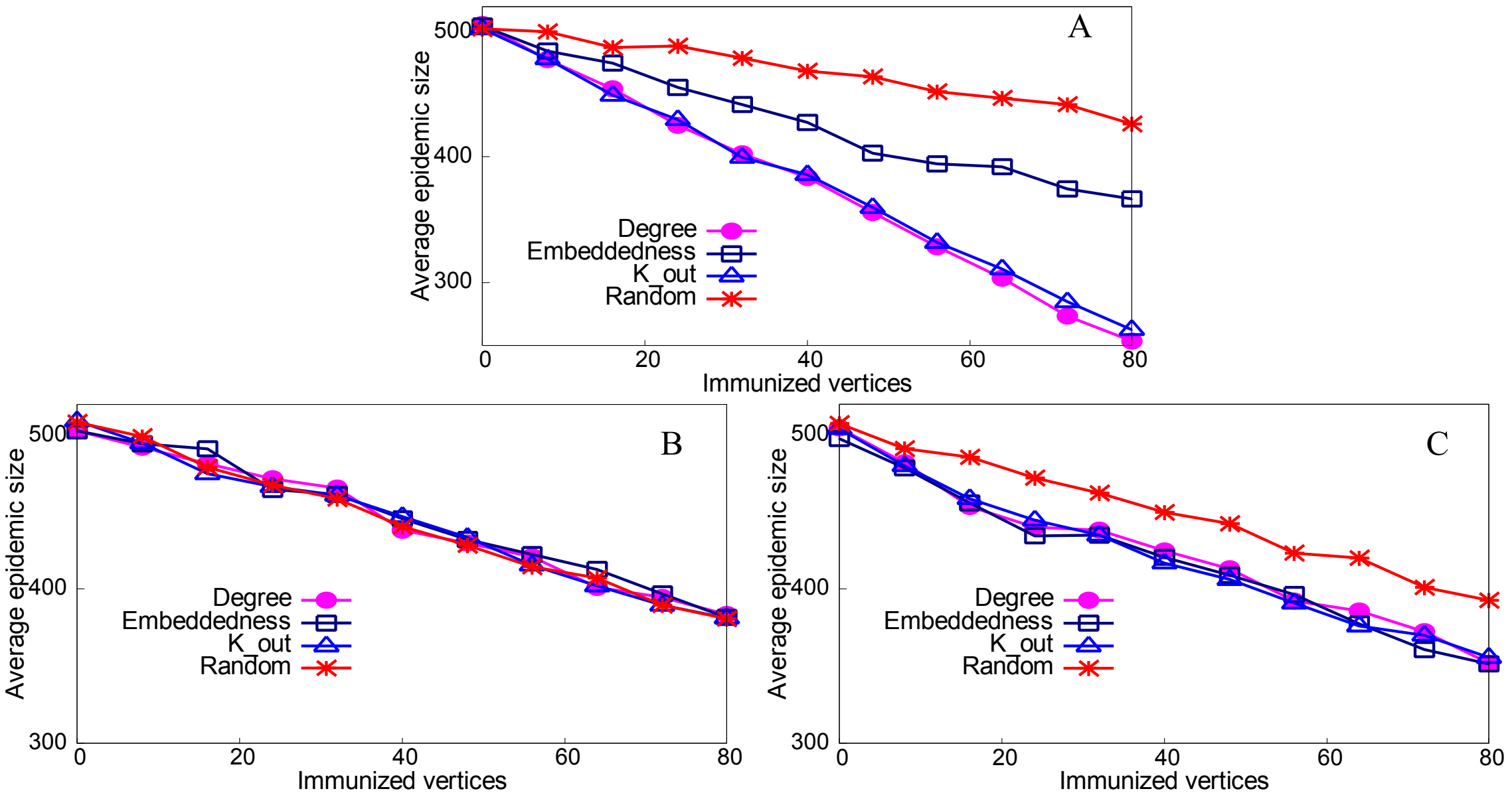

**Figure 13. Performance of four strategies on controlling disease in Email network.** Effect on epidemic size, varying number of immunized vertices. A. Immunizing top-ranked vertices in underlying network. B. Immunizing top-ranked vertices in reconstructed network: Random Path Method. C. Immunizing top-ranked vertices in reconstructed network: High-degree Path Method.

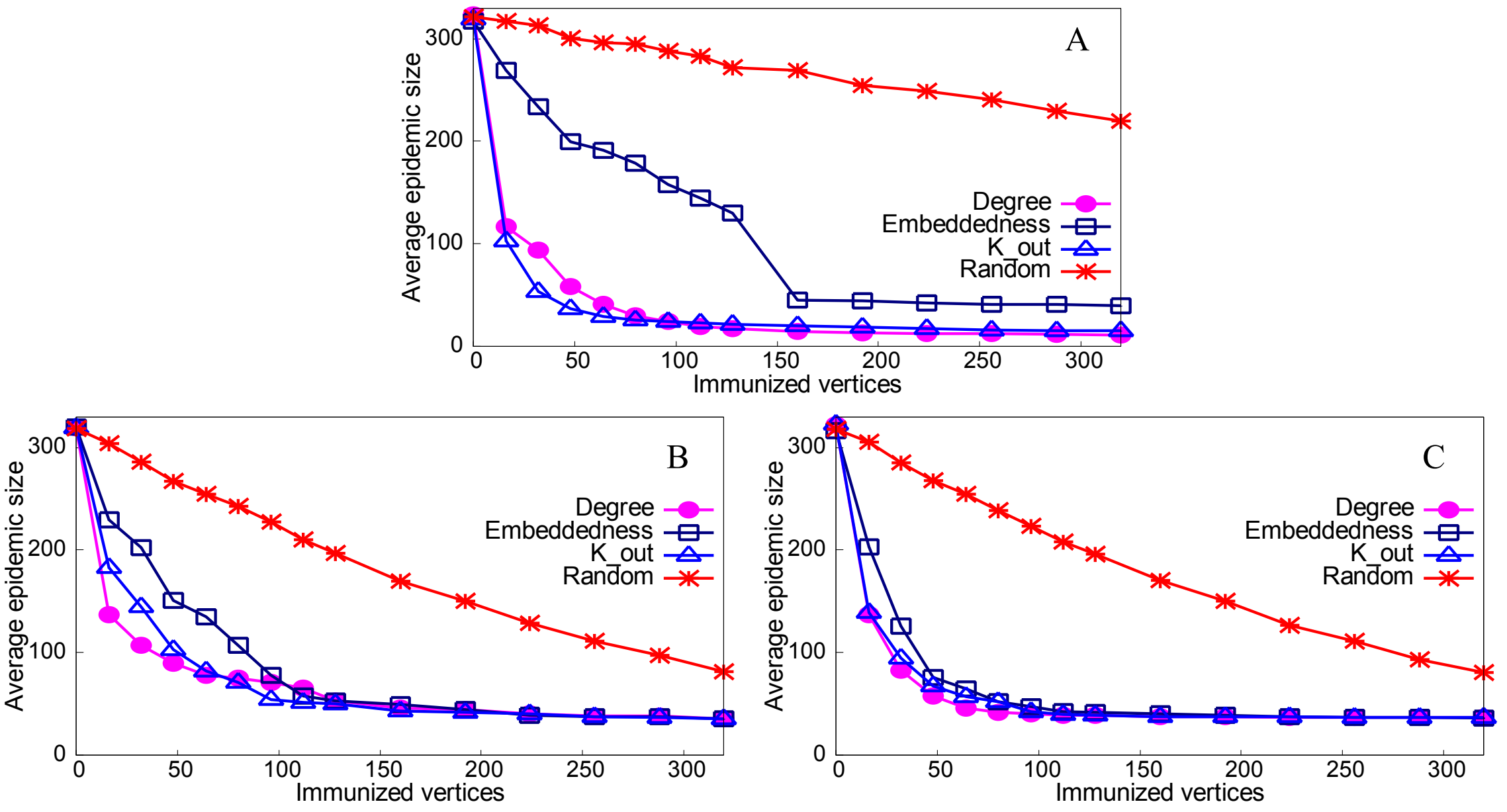

**Figure 14. Performance of four strategies on controlling disease in Blogs network.** Effect on epidemic size, varying number of immunized vertices. A. Immunizing top-ranked vertices in underlying network. B. Immunizing top-ranked vertices in reconstructed network: Random Path Method. C. Immunizing top-ranked vertices in reconstructed network: High-degree Path Method.

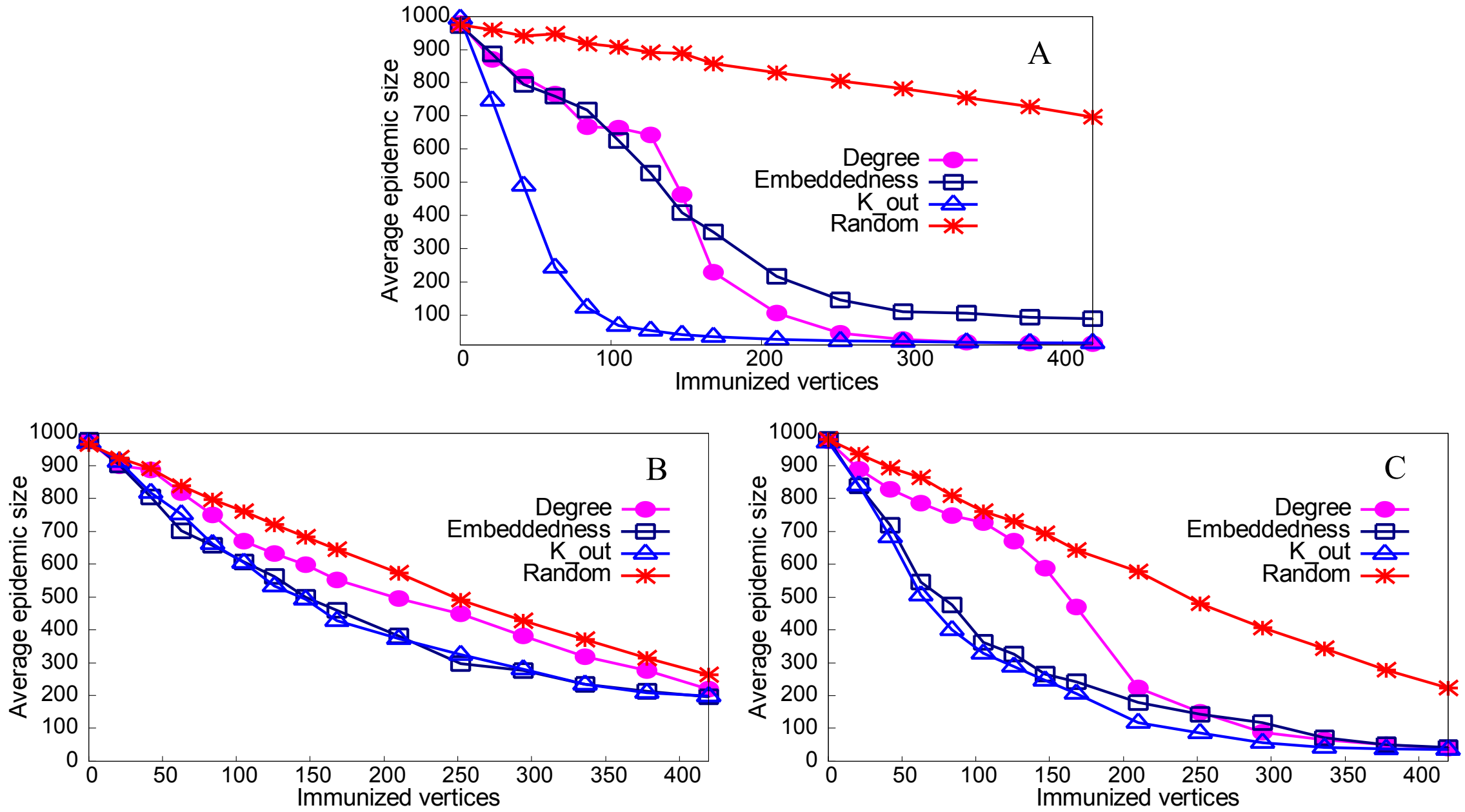

**Figure 15. Performance of four strategies on controlling disease in CA-GrQc network.** Effect on epidemic size, varying number of immunized vertices. A. Immunizing top-ranked vertices in underlying network. B. Immunizing top-ranked vertices in reconstructed network: Random Path Method. C. Immunizing top-ranked vertices in reconstructed network: High-degree Path Method.

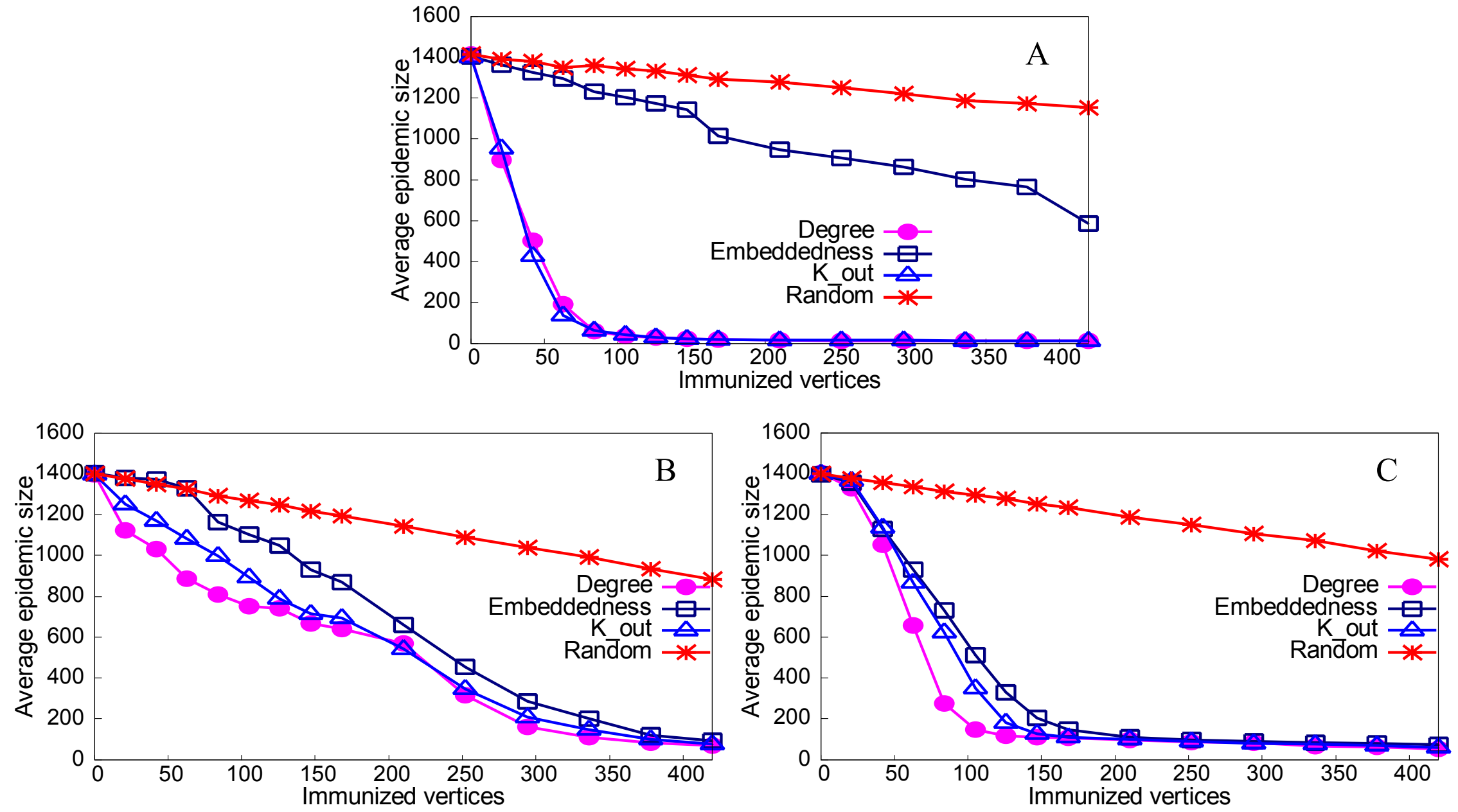

**Figure 16. Performance of four strategies on controlling disease in Erdös1997 network.** Effect on epidemic size, varying number of immunized vertices. A. Immunizing top-ranked vertices in underlying network. B. Immunizing top-ranked vertices in reconstructed network: Random Path Method. C. Immunizing top-ranked vertices in reconstructed network: High-degree Path Method.

In summary, immunizing a small number of key vertices in the reconstructed network can control the spread of infection almost as well as immunizing the same number of key vertices in the underlying network. The $k_{out}$ strategy is usually the best in the underlying network, but degree is the best strategy in the reconstructed network.

Finally, we investigate how sensitive our method is to the size of the sample. In Figure 17 we plot the effect on epidemic size in two of our real networks. We immunize the top 1% of vertices, according to degree and $k_{out}$, varying the size of the reconstructed network from which the vertices are chosen. The results suggest that our method is not particularly sensitive to the sample size, provided that the reconstructed network (size $n_t$) is at least 5% of the underlying network size. It is worth mentioning that, although the underlying network is generally unknown, its size may be easy to estimate (e.g., the number of drug users in a city can be estimated from the city's population and the average prevalence of drug use). Therefore, the sample size ($n_r$ and $n_f$) can be determined to make $n_t$ be approximately the desired fraction of the underlying network size.

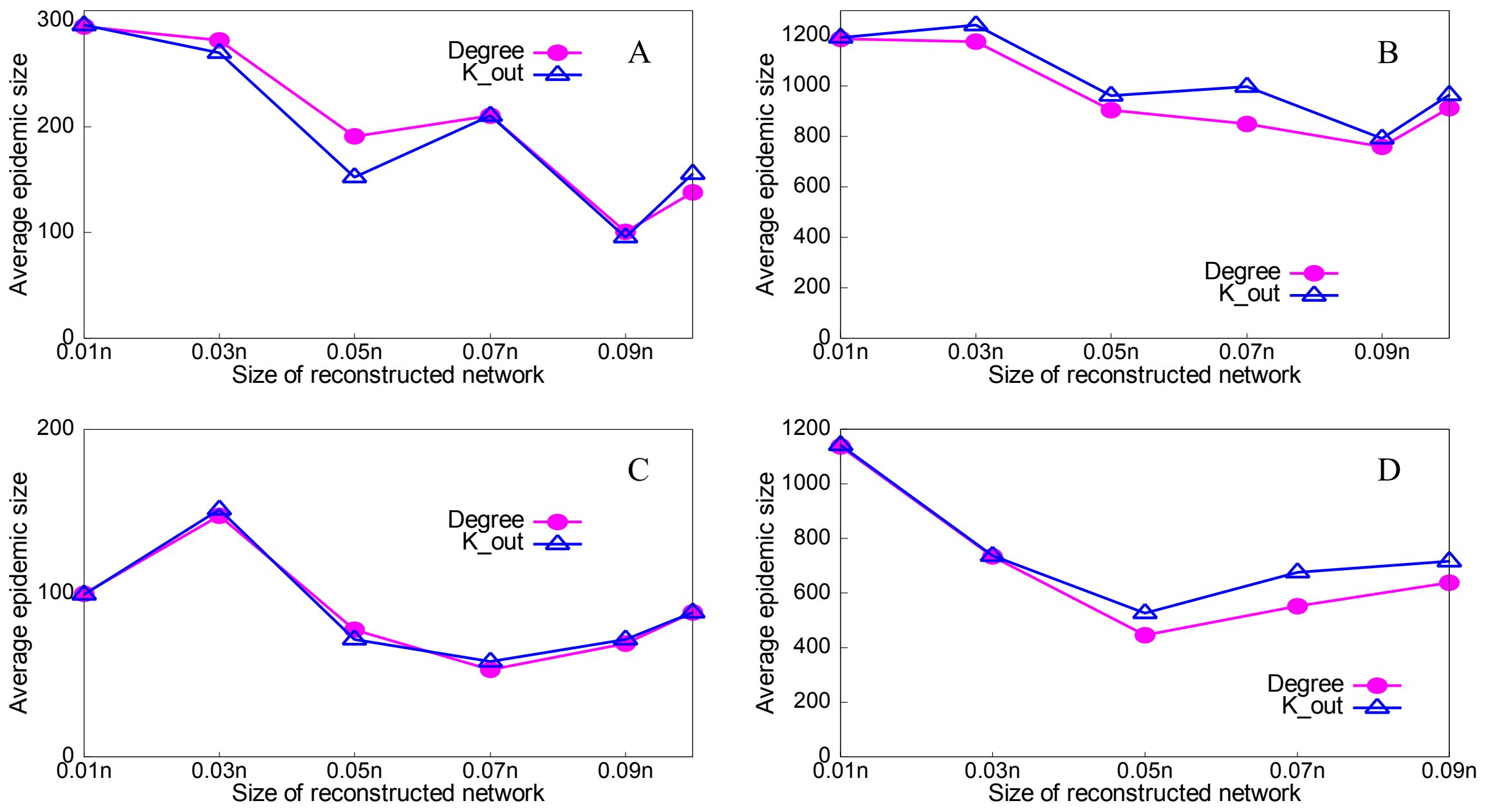

**Figure 17. Performance of degree and $k_{out}$ properties in controlling disease in real-world networks.** Effect on epidemic size, varying the size of the reconstructed network. (a) Blogs (RPM). (b) Erdös1997 (RPM). (c) Blogs (HPM). (d) Erdös1997 (HPM).

## Conclusions

We have described a sampling technique that allows some of the structure of networks to be recovered. Sampling strategies such as RDS [8] are already widely used in epidemiology to sample data from hidden populations. Indeed, Ref. [18] has examined the relation between community structure and RDS, showing that community structure in a network has an adverse effect on the sampling ability of RDS. However, RDS has not hitherto been used to *find* community structure in networks.

Our method does this by sampling trees and attempting to transform the trees to networks. The last step is a kind of missing data problem, in which the missing information is the identity of the pairs of vertices that represent the same individual. We solve this by using vertex attributes to calculate the probability that two vertices represent the same individual and coalesce them probabilistically. This is reminiscent of probabilistic linkage techniques, often used to combine data about the same individuals from multiple databases [19].

The results indicate that, not surprisingly, our method can reconstruct networks more accurately when descriptions are more precise. An interesting point is that community structure can be recovered quite well even when descriptions are not very accurate, because of the robustness of community detection methods. When the network features assortative mixing, which is common with community structure, the partitions found are even more accurate than in non-assortative networks. We also find that the ranking of vertices, with respect to degree and $k_{out}$, can be estimated quite accurately. Even though our estimate applies only to a small fraction of the contact network, it is remarkable that the spread of disease can be controlled well by immunizing the key vertices discovered in this way.

We have actually described two methods: the High-degree Path Method (HPM) and the Random Path Method (RPM). The HPM generally performs better. This is because the sampled vertices have higher degree and are more likely to form larger sample trees, and therefore larger components, in the reconstructed network; these components contain more information about the network structure than a larger number of very small components. However, the RPM also works well and is more practical because it is usually easier to discover *any* neighbors of a respondent than to find *all* of them, as the HPM requires.

It remains to be seen how well the technique will work in real life. This depends largely on how accurately respondents describe their friends, which will vary according to the nature of the survey. To evaluate the reconstructed network, we would also need to have access to the real underlying network, which is rarely available.

One area for future work is clearly to test the techniques in practice, by targeting real members of a clearly defined small population whose network structure can be verified. This is likely to be quite expensive because of the need to interview respondents. Another idea is to modify our sampling procedure to make it interactive: for example, if the probability of coalescing two friend vertices (friends of respondents $r_1$ and $r_2$) is too low, we could ask $r_1$ and $r_2$ to provide more detailed descriptions of those friends, or we could provide $r_1$'s friend description to $r_2$, for confirmation.

## Acknowledgements

We would like to thank Caroline Colijn, Katy Turner, Harriet Mills, and Nick Fyson for comments on earlier experiments.

## References

1. Lü L, Zhou T (2011) Link prediction in complex networks: a survey. Physica A 390: 1150-1170.
2. Leskovec J, Huttenlocher D, Kleinberg J (2010) Predicting positive and negative links in online social networks. In: WWW 2010: Proceedings of the 19th International Conference on World Wide Web. Raleigh, North Carolina, USA, pp 641-650.
3. Guo F, Yang Z, Zhou T (2012) Predicting link directions via a recursive subgraph-based ranking. Available: arXiv:1206.2199.
4. Miller JC (2009) Spread of infectious diseases through clustered populations. J. R. Soc. Interface 6: 1121-1134.
5. Salathé M, Jones JH (2010) Dynamics and control of diseases in networks with community structure. PLoS Comput. Biol. 6: e1000736.
6. Newman MEJ (2002) Assortative mixing in networks. Phys.Rev. Lett. 89: 208701.
7. Borgatti SP (2006) Identifying sets of key players in a network. Comput. Math. Organ. Theory 12: 21-34.
8. Heckathorn DD (1997) Respondent-driven sampling: a new approach to the study of hidden populations. Social Problems 44: 174-199.
9. Eames KTD, Keeling MJ (2003) Contact tracing and disease control. Proc. R. Soc. B-Biol. Sci. 270: 2565-2571.
10. Lancichinetti A, Fortunato S, Radicchi F (2008) Benchmark graphs for testing community detection algorithms. Phys. Rev. E 78: 046110.
11. Guimerà R, Danon L, Diaz-Guilera A, Giralt F, Arenas A (2003) Self-similar community structure in a network of human interactions. Phys. Rev. E 68: 065103.
12. Gregory S (2007) An algorithm to find overlapping community structure in networks. In: Proceedings of the 11th European Conference on Principles and Practice of Knowledge Discovery in Databases (PKDD 2007). Berlin, Germany: Springer-Verlag, pp 91–102.
13. Leskovec J, Kleinberg J, Faloutsos C (2007) Graph evolution: densification and shrinking diameters. ACM Trans. Knowl. Discov. Data 1: 2.
14. Batagelj V, Mrvar A (2000) Some analyses of Erdös collaboration graph. Soc. Networks 22: 173-186.
15. Rosvall M, Bergstrom CT (2008) Maps of random walks on complex networks reveal community structure. Proc. Natl. Acad. Sci. USA 105: 1118-1123.
16. Lancichinetti A, Kivelä M, Saramäki J, Fortunato S (2010) Characterizing the community structure of complex networks. PLoS ONE 5: e11976.
17. Mendenhall W, Scheaffer RL, Wackerly DD (2002) Mathematical Statistics with Applications. CA: Duxbury Press.
18. Goel S, Salganik MJ (2009) Respondent-driven sampling as Markov chain Monte Carlo. Stat. Med. 28: 2202-2229.
19. Jaro MA (1995) Probabilistic linkage of large public health data files. Stat. Med. 14: 491-498.